\documentstyle[psfig]{l-aa}
\begin{document}
\thesaurus{11.13.1, 13.18.1, 04.19.1, 09.08.1, 03.13.6}
\title{Dissipation of Magnetic Fields in the Galactic Halo}
\subtitle{\normalsize \it Dedicated to Professor Karl Schindler on the occasion of his 65th birthday }
%\thanks {}
%
%
%
\author{F. Zimmer\inst{1}, H. Lesch\inst{2}, G.T. Birk\inst{2}
}
\offprints{F. Zimmer}
\institute{
   Radioastronomisches Institut der Universit{\"a}t Bonn,
   Auf dem H{\"u}gel 71, D-53121 Bonn, Germany
\and
   Institut f{\"u}r Astronomie und Astrophysik der Universit{\"at} M{\"u}nchen,
   Scheinerstra{\ss}e 1, D-81679 M{\"u}nchen, Germany
}
\date{Received June 27, accepted October 4, 1996}
\maketitle
\begin{abstract}
This paper shows that magnetic reconnection could be an important
heating process in cosmic gases.
In any volume where magnetized plasmas
collide, the dissipation of magnetic energy via reconnection seems to be
unavoidable. Since most
cosmic plasmas are highly conductive, the magnetic field lines are transported
with the gas and no dissipation occurs for the most part of the volume.
This ideal frozen-in property  of the magnetic field is broken in small volumes if
field gradients with different field polarity appear, in which
localized dissipative effects, e.g. anomalous resistivity, become important.
On the base of X-ray measurements
exhibiting a clear connection of infalling high-velocity clouds (HVC) with ROSAT "hotspots"
we perform resistive magnetohydrodynamic simulations to investigate the capabilities
of magnetic dissipation as a major heating process in the interaction zone of the cloud with the
halo.
The main result is that
in the physical environment of a galactic halo
heating by externally driven magnetic reconnection cannot be suppressed by
thermal conduction and/or radiative cooling. Thus, the gas reaches the maximum temperature
given by the magnetic field pressure in the interaction zone of the HVC with the galactic halo.

\end{abstract}
\section{Introduction}

Radio observations clearly reveal that
the halo of the Milky Way is a magnetized plasma
(Haslam et al. 1982). The radio emission is due to
nonthermal synchrotron radiation of relativistic electrons gyrating
around a magnetic field with an average field strength of several $\mu$G (Beuermann, Kanbach and Berkhuijsen 1985).
The relativistic particles move through a fully or partially ionized
plasma, which consists of several gas components, starting from HI with temperatures
of some 100 K and extending up to X-ray emitting gas with temperatures of
about a million K (e.g. Lesch et al. 1996 for a recent conference about the
galactic halo). The halo gas is agitated by the activity in the disk (supernova remnants, stellar winds,
bubbles and superbubbles, etc.), which led to the picture of Galactic 
Fountain (e.g. Kahn 1981). This fountain is formed by hot gas rising from the
galactic disk. At some height the gas starts to cool and may
partially fall back onto the disk in form of high-velocity clouds (Kahn 1991).
Thus, the halo gas may also be agitated by the recurrent high-velocity clouds.
Such HVCs have been observed long before the scenario of galactic fountains
was developed (Muller et al. 1963) and it was anticipated that they interact
with the halo gas or the disk itself (Oort 1970). One should expect that
the interaction should reveal itself by some enhanced radiation originating
from heated gas.

Hirth et al. (1985) first announced the detection of a HVC associated with soft X-ray
emission close to the Draco cloud, which itself may be interpreted as the result of a HVC-disk
interaction (Mebold et al. 1989). Since their data covered only the $ \frac{1}{4} $ keV energy range which is heavily
disturbed by absorption of neutral matter distributed along the line of sight,
their detection was not completely convincing.

Kerp et al. (1994) found clear evidence for the association of the HVC with $ \frac{3}{4} $
keV X-ray emission, which cannot be significantly attenuated by absorption. They
also detected a detailed anticorrelation of HVC and X-ray emission.
In a recent paper Kerp et al. (1996) were able to show that large parts
of the HVC complex C are associated with soft X-ray radiation. With the
new Leiden-Dwingeloo HI-survey (Hartmann and Burton 1995) they could
prove that the HVC velocity regime is connected with low-velocity atomic hydrogen
of the galactic disk by velocity bridges. In other words the impacting HVC gas
is dragged by the halo. This interpretation is corroborated by positional
correlation of the HI velocity bridges with the soft X-ray enhancement.
Thus, the edges of the HVC are sources of strong X-ray emission, revealing
a gas temperature of a few million degrees.

The striking conclusion that an impacting HVC is responsible for
the local X-ray emission raises the question about the physical mechanism
which releases the observed luminosity of about $10^{32}$ erg\,s$^{-1}$. An inelastic
collision of the HVC with the galactic disk converts kinetic energy into thermal
energy (Hirth et al. 1985). But this ram pressure process is unlikely the source of the X-rays
even of the $ \frac{3}{4} $ keV emission, because most of the HVC models predict
velocities that are approximately equal in all three spatial directions and of about
100 km\,s$^{-1}$ (Mebold et al. 1991). Therefore, the complete kinetic energy of the cloud
has to be transformed into heat in order to reach the observed few million K.

In Fig. 1 we sketch the scenario we have in mind. A HVC hits the Reynolds layer
(Reynolds 1991), which consists of an ionized, magnetized hydrogen plasma.
We assume that the magnetic field in the halo is preferentially directed parallel
to the disk, which usually is the case for spiral galaxies (Dumke et al. 1995).

\ \\
\vspace*{-1.5cm}
\ \\
\begin{figure}[ht]
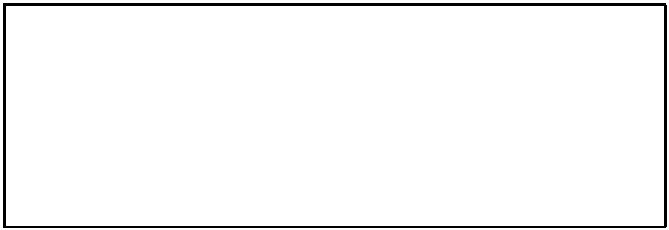

%\hspace*{1.cm}
%\vspace*{-2.5cm}
%\psfig{figure=fig1.ps,width=6cm,angle=270}
\picplace{2.95cm}
%\\[2.0cm]
\caption[GALAXY]{{\footnotesize{A HVC collides 
with the Reynolds layer, an ionized hydrogen layer of the galactic
halo which includes magnetic fields}}}
\end{figure}
\ \\
It is the aim of this contribution to investigate the probable braking mechanisms for the
HVC. Even though the impact velocities are of the order of the sound speed for a
gas of one million degree Kelvin, shock heating does not provide a sufficiently efficient
mechanism for X-ray temperatures (Zimmer et al. 1996).
In Section 2 we introduce magnetic reconnection as a major heating process in the interstellar
medium. The interaction of HVCs with the galactic disk can be considered as a prime example
for this process. Section 3 contains the details
of the resistive magnetohydrodynamic simulations of the cloud impact onto the galactic disk.
Several cooling processes competing with the magnetic heating are considered
in Section 4. Finally, we sum up our findings in Section 5.

\section {Magnetic dissipation}

It is well known that magnetic fields contribute significantly to the
interstellar pressure and energy reservoir. Most of the interstellar plasma
is highly conductive, i.e. we can write Ohm's law 
${\bf E} + \left( \frac{1}{c}{\bf v \times B} \right) = 0$
(if we neglect the Hall term, electron pressure term and inertia terms
for simplicity).
In this case the magnetic field is ``frozen'' in the plasma and transported
by plasma flows.
This property of the magnetic field changes drastically, e.g. 
if field lines with different directions approach each other and localized
dissipative regions form. The chain of processes which is triggered
then is called {\bf magnetic reconnection} (e.g. Priest 1985;
Biskamp 1994 and references therein).

Magnetic reconnection is a fundamental intrinsic property of 
agitated magnetized
plasmas with a non-zero electric field component along the magnetic field
(e.g. Schindler et al. 1991). Whenever, magnetic fields with
different field directions encounter, the
magnetic energy can partly be dissipated, either by accelerating particles or by
plasma heating. The approaching field lines correspond to parallel currents
which attract each other. Outside the forming current sheet the motion 
of the plasma
is "frozen-in", i.e. the magnetic field lines follow the plasma motion as if
they were frozen into it, which is provided by the high electrical conductivity.
The oppositely directed field lines following the plasma motion approach,
the field gradient steepens and the current density ${c\over{4\pi}}{\bf \nabla\times B}$
increases until strong dissipation sets in. 
Different from mere diffusion that occurs on a time scale given
by $\tau_D = L_o^2/\eta$ reconnection is a rather fast localized
dissipative process involving global changes in magnetic field topology.
The problem of reconnection is
to know how the dissipation of the currents with the density $j=e n_{\rm e} v_{\rm d}$
is provided ($n_{\rm e}$ is the electron number density per ${\rm cm^3}$ and
$v_{\rm d}$ denotes the drift velocity of the electrons relative to the protons.)

Since magnetic reconnection seems to be unavoidable in plasmas
with randomly mixed magnetic fields 
we propose that the encounter of a high velocity plasma with
the ``magnetic atmosphere'' of the Milky Way transfers the
kinetic energy of the clouds into heat via compressed, strained and teared 
magnetic field structures, which heavily dissipate the stored magnetic energy
in current sheets via magnetic reconnection.

Since magnetic reconnection corresponds to the dissipation of electric currents
the dissipation (or heating) rate $Q$ (in erg\,cm$^{-3}$\,s$^{-1}$) is

\begin{equation}
Q={j^2\over \sigma}.
\end{equation}
Dissipation is equivalent to either increased current density and/or
reduced electrical conductivity $\sigma$

\begin{equation}
\sigma={\omega_{\rm pe}^2\over{4\pi \nu_{\rm coll}}}.
\end{equation}

$\nu_{\rm coll}$ is the collision frequency and $\omega_{\rm pe}\sim 5.6\cdot 10^4
\sqrt{n_{\rm e}}$ is the electron plasma frequency.

We first give a qualitative estimate of
kinetic energy conversion into heat via magnetic reconnection and
rough estimates of the heating process. The details of the plasma processes involved
have been discussed by Lesch (1991).

The collision of a high velocity gas with the magnetized galactic disk can be
described in two ways: as an interaction of two magnetized plasmas, if the
HVC contains a magnetic field or otherwise as the interaction of
a non-magnetized plasma with a magnetized one, if only the disk is magnetized.

In the first case, the appearance of magnetic reconnection is obvious: the
magnetic field lines of the cloud and the field lines of the disk are compressed and randomly
mixed in a boundary layer. There is observational evidence that HVCs indeed
possess magnetic fields. Kaz\`es et al. (1991) detected a field strength of
about 11${\rm \mu}$G in a HVC via Zeeman measurements.

Pietz et al. (1996) detected HI-velocity-bridges which demonstrate that the clouds velocities
continuously decrease with decreasing distance to the disk, i.e. the clouds are slowed down.
The size of these bridges indicates that the interaction region
between the cloud and the halo has an extension of about 10$'$ or 4 pc assuming a distance of 1.5 kpc.
Using the Westerbork interferometer Wakker (1991) observed the substructure
of HVCs.
He found thin dense filaments with an extension of less than 0.4 pc assuming a distance
of 1.5 kpc again. These
filaments reveal turbulent motions inside the boundary layer down to little scales.
Kerp \& G{\"u}sten (1996) confirm this result via Zeeman measurements: They derive a magnetic
field strength of 20 - 30 $\mu$G, which changes on small scales inside the cloud.

Thus, we can use about
10 ${\rm \mu}$G as the value for the unperturbed field strength.
If such a magnetized cloud approaches the galactic disk the undistorted field
will be compressed until reconnection sets in due to localized dissipative
regions, since for most of its path the
magnetic field is frozen into the cloud. Such an encounter of two magnetized plasmas
could be envisaged as the collision of two magnetized networks. As long as
both magnetic field structures collide with some angle reconnection will
be driven by the plasma motion of the cloud. A huge number of current sheets
will be built up in the whole interaction volume, which are accompanied by
dissipation via plasma heating (Lesch and Bender 1990).

In the second case, the magnetic field of the disk will be distorted by the shear flow
in the boundary layer between the incoming HVC and the disk.
 As long as the field is frozen-into
the plasma motion the field is mixed and again field lines with different directions
will encounter and the kinetic energy of the cloud is transferred into heat
via the formation current sheets and Ohmic dissipation.

The physical scenario we have in mind is sketched in Fig. 2.
Assuming a distance of the clouds of about 1.5 kpc the HVC interacts with the Reynolds layer, the
ionized hydrogen gas layer of the galactic halo. The clouds themselves are ionized by
cosmic rays and the soft-X-ray-background as well.
The impact of the cloud into the Reynolds layer is shown in the left upper
figure which represents the most interesting part of the ROSAT-image schematically.

\begin{figure}[ht]
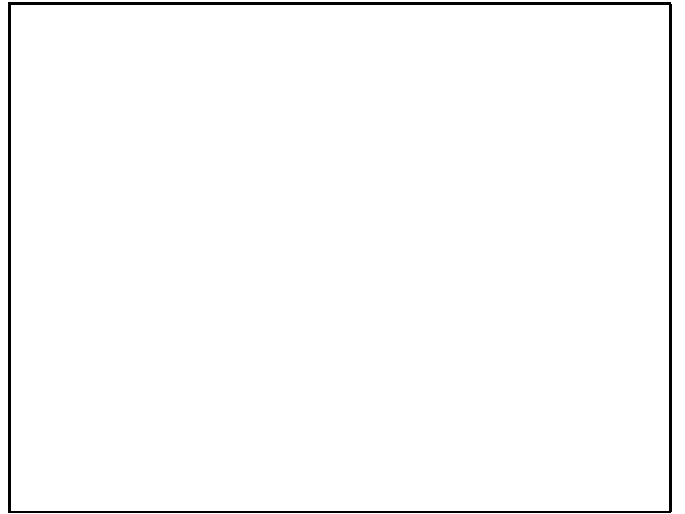

%\hspace*{0.75cm}
%\vspace*{-0.5cm}
%\psfig{figure=fig2.ps,width=7cm}
%\vspace*{0.25cm}
\picplace{6.75cm}
\caption[SCENARIO]{{\footnotesize{The screnario we have in mind: A HVC interacts with
the Reynolds layer and a large number of small scale reconnection sheets is formed.}}}
\end{figure}

For both cases the maximum temperature can be estimated by considering
the equilibrium between the kinetic energy density and magnetic energy density
or thermal energy density, respectively:
\begin{equation}
{1\over 2} n_{\rm H}^{\rm HVC} m_{\rm p} v_{\rm HVC}^2 \simeq {B_{\rm 0}^2\over{8\pi}}
\simeq n_{\rm e}^{BL} k_{\rm B} T_{\rm e}
\end{equation}
here $v_{\rm HVC}$ is the velocity of the HVC,
$n_{\rm H}^{\rm HVC}$ is the neutral density of the HVC and $n_{\rm e}^{\rm BL}$
is the electron density in the boundary layer. $T_{\rm e}$ denotes the electron
temperature, $k_{\rm B}$ is the Boltzmann constant and $m_{\rm p}$ is the proton mass.

The resulting temperature is given by
\begin{equation}
T_{\rm e}\simeq {n_{\rm H}^{\rm HVC}\over{n_{\rm e}^{\rm BL}}}
{1\over k_B}{1\over 2} m_{\rm p} v_{\rm HVC}^2
\end{equation}

which gives
\begin{equation}
T_{\rm e}\simeq 6\cdot 10^6 {\rm K}
\left[{v_{\rm HVC}\over{100\, km s^{-1}}}\right]^2
\left[{n_{\rm H}^{\rm HVC}/n_{\rm e}^{\rm BL}\over {10}} \right]
\end{equation}
We note that the Eq. (5) is an underestimate of the heat transfer since the
density contrast $n_{\rm H}^{\rm HVC}/n_{\rm e}^{\rm BL}$ is larger
than 10 (Wakker 1991).

The attainable temperatures are definitely high enough
to explain the edge brightening of the HVC in the X-ray range.

The global energy budget can be estimated as follows: the energy density of the
magnetic field $U=B^2/8\pi$ is dissipated with an efficiency $\epsilon$ across an area
A of about (4 pc)$^2$ (taken from the observations). For the velocity
with which energy is converted during the reconnection process ($v_{diss}$)
we use a tenth of the velocity of the HVC $v_{\rm diss}\sim$ 10 km\,s$^{-1}$.
This results in a total luminosity

\begin{eqnarray}
L_{\rm rec} &=& U\, \epsilon\, v_{\rm diss} \nonumber \\
&\sim& 10^{33}\, \frac{erg}{s}
\left[{\epsilon\over{0.1}}\right]
\left[{B\over{10\, \mu G}}\right]^2
\left[{v_{\rm diss}\over{10\, \frac{km}{s}}}\right]
\left[{A\over{(4\, pc)^2}}\right]
\end{eqnarray}

After these qualitative estimates we present in the next section our numerical
resistive magnetohydrodynamical simulations.

\section{Simulation Results}
We perform resistive magnetohydrodynamic simulations (for details
concerning the code see Otto 1990) to investigate
the impact of a magnetized gas cloud onto a magnetized gas.
We numerically integrate the set of basic equations:

\ \\
\fbox{\parbox{8.65cm}{
\begin{itemize}
\item[~] {\bf \hspace*{0.5cm} \hspace*{-0.75cm} Equation of Continuity}:
      \[\hspace*{0.5cm} \hspace*{-0.75cm}\frac{\partial \rho}{\partial t} = - \nabla \cdot (\rho \vec{v} )\]
\item[~] {\bf \hspace*{0.5cm} \hspace*{-0.75cm} Momentum Equation}:
      \[\hspace*{0.5cm} \hspace*{-0.75cm} \frac{\partial }{\partial t} (\rho \vec{v}) =  - \nabla \cdot (\rho \vec{v} \circ \vec{v} ) - \nabla p + \, \frac{1}{4 \pi} \left( \nabla \times \vec{B} \right) \times \vec{B}  \]
\item[~] {\bf \hspace*{0.5cm} \hspace*{-0.75cm} Energy Equation}:
      \[\hspace*{0.5cm} \hspace*{-0.75cm}\frac{\partial p}{\partial t} = - \vec{v} \cdot \nabla p \,- \, \gamma p \,\nabla \cdot \vec{v} + (\gamma-1)\,\eta \left(\frac{c}{4 \pi} \right)^2 (\nabla \times \vec{B}\,)^2\]
\item[~] {\bf \hspace*{0.5cm} \hspace*{-0.75cm} Induction Equation}:
      \[\hspace*{0.5cm} \hspace*{-0.75cm}\frac{\partial \vec{B}}{\partial t} = \nabla \times ( \vec{v} \times \vec{B} ) - \frac{c^2}{4 \pi} \nabla \times \left( \eta \,\nabla \times \vec{B} \right) \]
\end{itemize}
}}

\ \\
\ \\
Here $\rho$, $p$, $\vec{v}$, $\vec{B}$ and $\eta$ denote 
the mass density, thermal pressure,
plasma velocity, the magnetic field and the resistivity.
All quantities are made dimensionless by a normalization to typical
parameters. Length scales are normalized to a
typical length $L_o$, $\rho$ is normalized to a mass density 
$\rho_o = m_o\,n_o$ (where
we take $m_o$ to be the ion mass), ${B}$ to a typical magnetic field strength $B_o$, $p$ to $p_o = \frac{B_o^2}{8 \pi}$ and the
plasma velocity to the Alfv\'en velocity $v_A = \frac{B_o}{\sqrt{4 \pi \rho_o}} $. The time scale is
normalized to the Alfv\'en transit time $\tau_A = \frac{L_o}{v_A} $.

\ \\
Here we present some results of 2$\frac{1}{2}$D simulations
($\partial / \partial z = 0$). The resistive
MHD-equations above are integrated on a 2D domain shown in Fig. 3.

\ \\
\begin{figure}[ht]
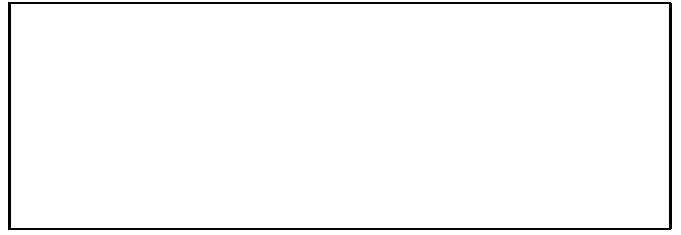

%\hspace*{0.85cm}
%\vspace*{0.2cm}
%\psfig{figure=fig3.ps,width=7cm}
%\vspace*{-0.1cm}
\picplace{3cm}
\caption[NUMERICALAREA]{{\footnotesize{The MHD-equations are integrated on the 2D area,
shown as a dashed plane.}}}
\end{figure}
\\[-0.5cm]

Initially the magnetic field has the following
form:
\[ \vec{B} = \vec{B}_i \, tanh\left(\zeta \left[ y-\frac{y_{max}+y_{min}}{2}\right]\right)\,\vec{e}_x \]
and thus, the current density reads:
\[ \vec{j} = \vec{j}_i \, sech^2\left(\zeta \left[ y-\frac{y_{max}+y_{min}}{2} \right]\right)\,\vec{e}_z \]

During the dynamical evolution magnetic flux is transported into 
the numerical domain, 
and magnetic field lines of different polarity are
compressed. 
Whenever magnetic fields with different polarity approach, a current sheet is formed. Outside
this sheet the motion of the plasma is ''frozen in'', that means, the magnetic field lines follow the plasma
motion as if it were frozen into it. This is due to the high electric conductivity in the plasma outside
the current sheet. So, when the plasmas collide, the magnetic field lines follow this motion, the
field gradient steepens and the current density $ j \sim \nabla \times \vec{B} $ increases significantly.
On the other hand $j$ is proportional to the drift velocity of the electrons $ j \sim n e v_D $, so that $ v_D $
increases with $j$.

The growth of the current density, i.e. of the drift velocity is limited by
microscopic plasma instabilities (e.g. Sagdeev 1979). If the drift velocity
exceeds the thermal velocity of the particles, several wave modes that
are excited by pressure gradients or electric currents can grow
unstable. The nonlinear saturation phase of these microinstabilities
gives rise to  electrical resistance mediated by
wave-particle interactions. Momentum transfer between the charges species
is caused via turbulent electromagnetic fields, i.e.
the classical resistivity changes into anomalous resistivity. 
Since the anomalous collision frequency
is considerably larger than the classical Coulomb value, the onset
of microturbulence increases the dissipation rate 
by many orders of magnitude.

We model the effect of anomalous resistivity by:

\[ \eta = \eta_o \cdot | \vec{j} - \vec{j}_c| \cdot \frac{1}{cosh(\xi\,x)} \]
if the current density $ | \vec{j} | $ exceeds the critical value $ j_c$
($\xi$ measures the scale length on which the resistivity vanishes
in the $x$-direction ($\frac{1}{\xi} \approx 7\cdot 10^9$ cm)). 

Otherwise the resistivity vanishes. 

\onecolumn
\begin{figure}[ht]
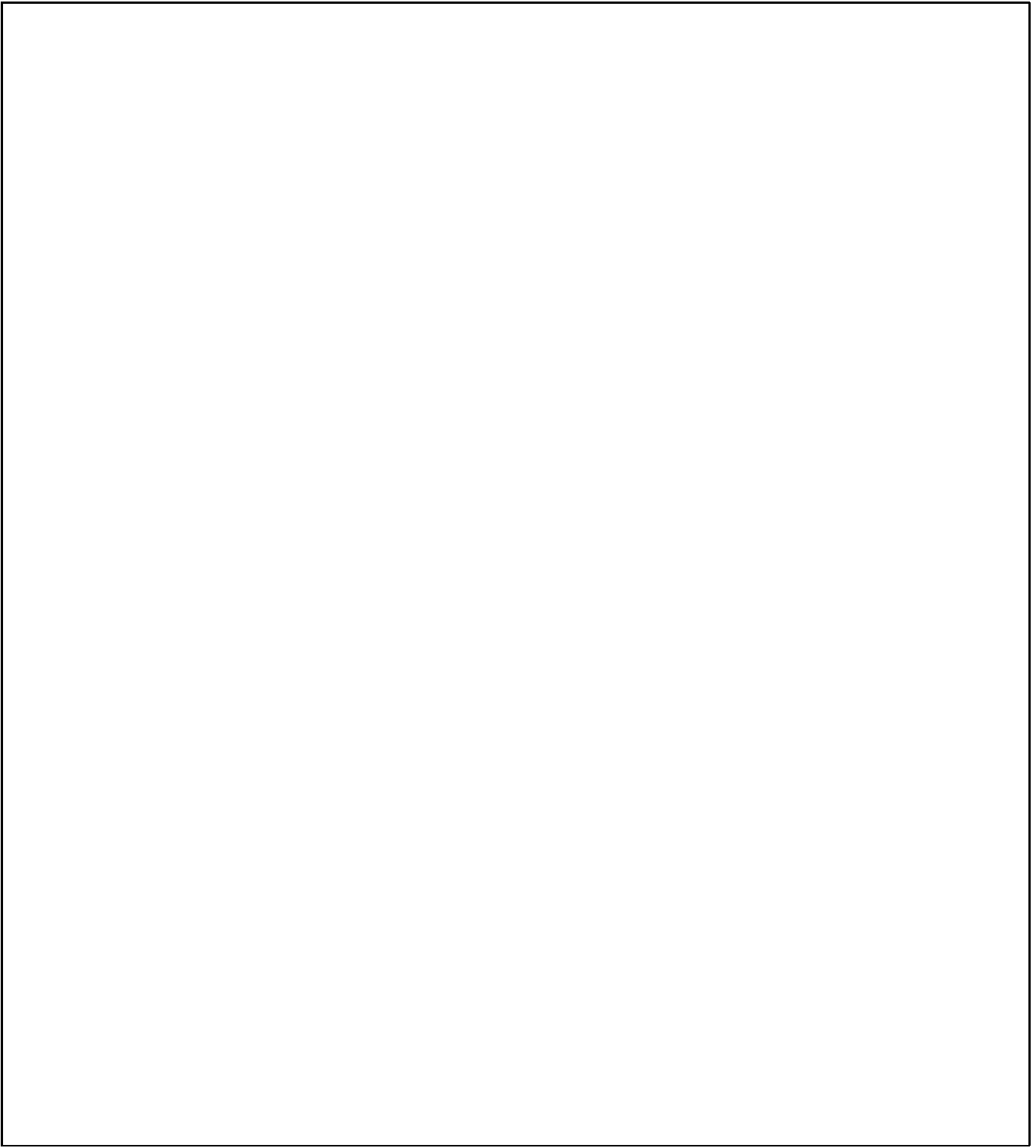

%\ \\
%\\[-1.0cm]
%\hspace*{-2.0cm}
%\psfig{figure=fig4.ps,width=20cm}
%\\[-5.25cm]
\picplace{20cm}
\caption[DENSITY]{{\footnotesize{The time development of the number density n with an initial
temperature of 9$\cdot$10$^5$ K and an initial number density of 0.01 particles per cm$^3$. The typical
length scale $L_o$ is about 1.5 $\cdot$ 10$^9$ cm.}}}
\end{figure}
\twocolumn

The drift velocity is associated with the magnetic field gradient via
\begin{equation}
v_{\rm D}\simeq{c\over{4\pi e n_{\rm e}}}{B\over l}\,
\end{equation}
Eq. (7) means that the larger the magnetic field gradient is the higher is the drift
velocity. So one can expect that in the innermost current sheet 
plasma waves are excited. We note that in general one has to
distinguish between magnetized and unmagnetized instabilities. The
former could not appear in the neutral sheet (with $B=0$). 
Thus, these instabilities do not directly produce anomalous resistivity 
in magnetic null regions (details see Lesch 1991).

In what follows we consider unstable lower hybrid drift waves
as the mediator for anomalous
resistivity in the reconnection region. Such waves are excited by
local pressure gradients and grow unstable if the drift velocity
associated with these gradients exceeds a critical value.
lower hybrid drift instability (LHI).
Two properties in favor
of this instability are (1) the mode can be excited in relatively broad current
sheets since the necessary drift velocity is the thermal velocity of the ions
$l\simeq (\frac{m_i}{m_e})^{1/4} r_{\rm Li}$ (where $r_{\rm Li}$ is the ion Larmor
radius) and (2) the mode is insensitive to the temperature ratio $\frac{T_e}{T_i}$.
The anomalous collision frequency associated with LHI is approximately equal to
the lower-hybrid frequency $\omega_{\rm LH}$ (e.g. Shapiro et al. 1994, Sotnikov et al. 1978)
\begin{equation}
\nu_{\rm coll}^{LHI}\simeq \omega_{\rm LH}\simeq 4\times 10^5\, B\, s^{-1}.
\end{equation}

Therefore, in the following we assume
$\nu_{\rm coll}\simeq \omega_{\rm LH}$
and $v_{\rm D} \simeq v_{thi}$.

\ \\
The first example shows a situation, where the initial temperature is about \mbox{900\,000 K}
and the initial density is about 0.01 particles per cm$^3$. The initial pressure
and initial density are constant. The ``global'' Alfv\'en speed is about $ 3 \cdot 10^7$ cm\,s$^{-1}$ ;
the magnetic field strength is about 17 $\mu$G. 

Fig. 4 shows the time development of the number density. The time is given
in Alfv\'en times, one Alfv\'en time is about 50 sec (for comparison
the diffusion time is some 10$^4$ sec).
We see that the plasma
is compressed by the magnetic field and the density grows very quickly.
\ \\
\begin{figure}[ht]
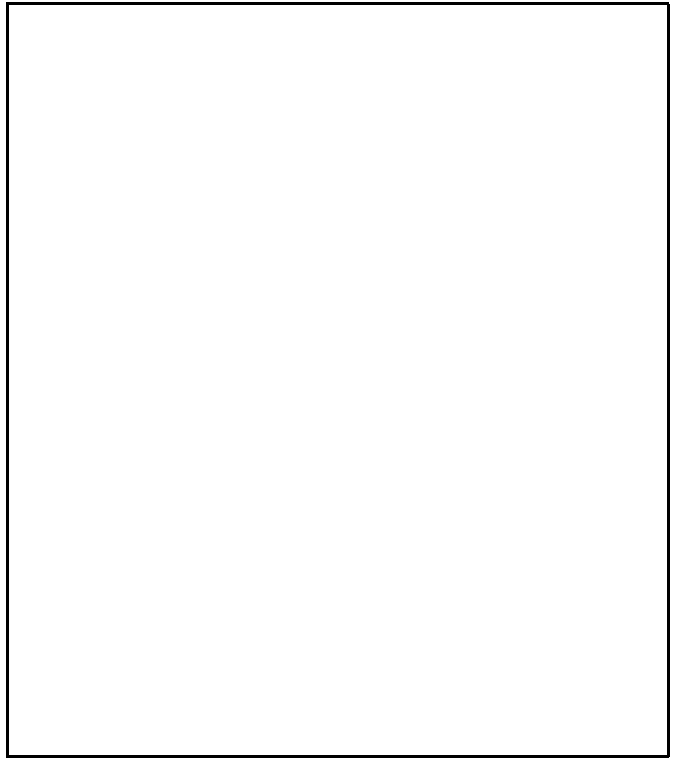

%\vspace*{-1.5cm}
%\psfig{figure=fig5.ps,width=8cm}
%\vspace*{-0.75cm}
\picplace{10cm}
\caption[OUTFLOW]{{\footnotesize{The outflow component of the velocity after 75 Alfv\'en times}}}
\end{figure}
\\[-0.5cm]

Further we see that the
density then becomes smaller again inside the reconnection zone. This is because pressure forces and the Lorentz force
accelerates the plasma out of the zone. This can be seen in Fig. 5 showing 
the outflow component of the plasma velocity after 75 Alfv\'en times,
which reaches about 50 - 60 \% of the ''global'' Alfv\'en speed.
The time-development of the magnetic field is illustrated in Fig 6.
Magnetic field lines of opposite polarity encounter, the field lines 
reconnect and the Lorentz forces arising in these configurations
accelerate the plasma out of the reconnection zone.

\onecolumn
\begin{figure}[ht]
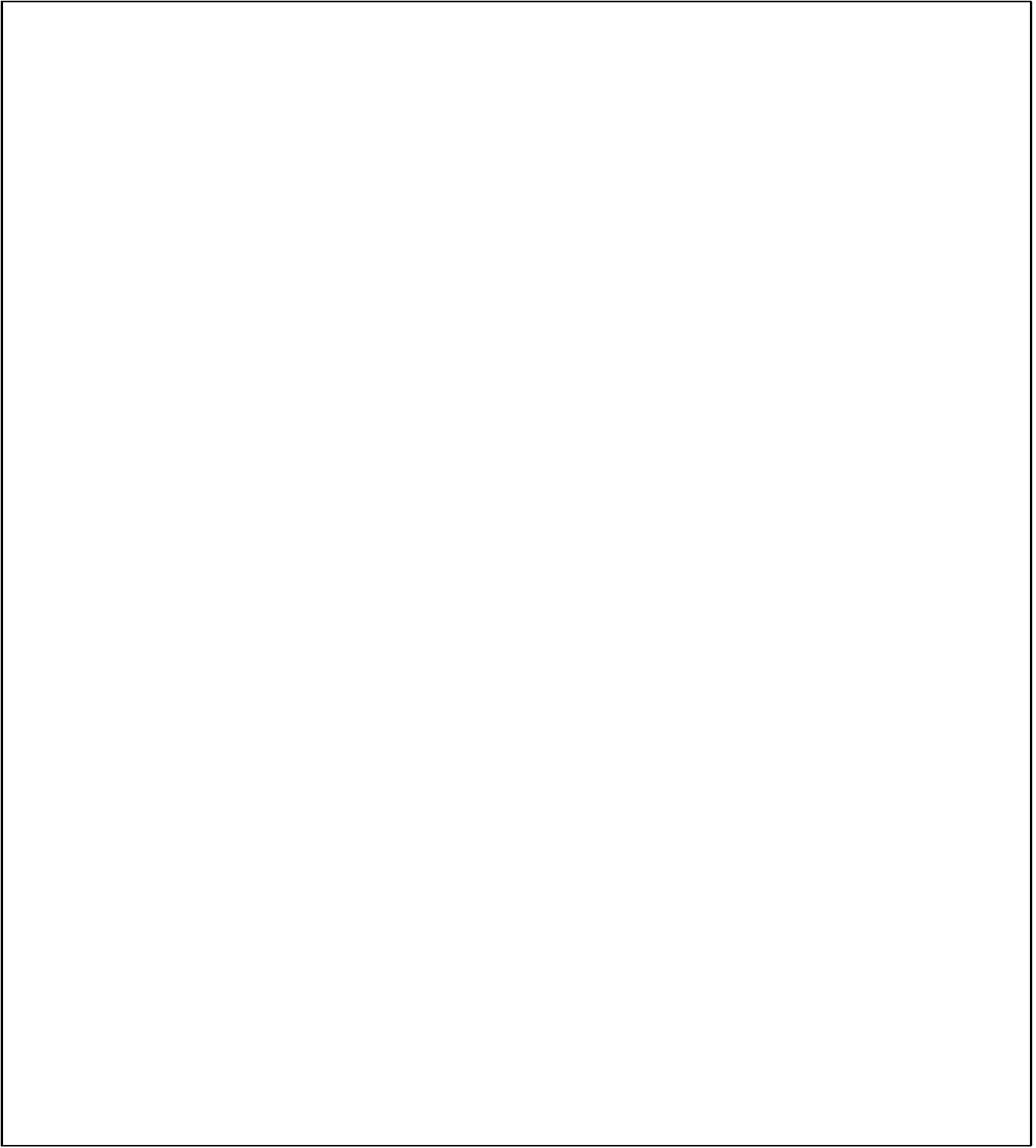

%\ \\
%\\[-1.0cm]
%\hspace*{0.0cm}
%\psfig{figure=fig6.ps,width=20cm}
%\\[-5.25cm]
\picplace{20cm}
\caption[MAGNETICFIELD]{{\footnotesize{Magnetic field lines of different polarity flow together
and they reconnect. ``Magnetic islands'' form.}}}
\end{figure}
\begin{figure}[ht]
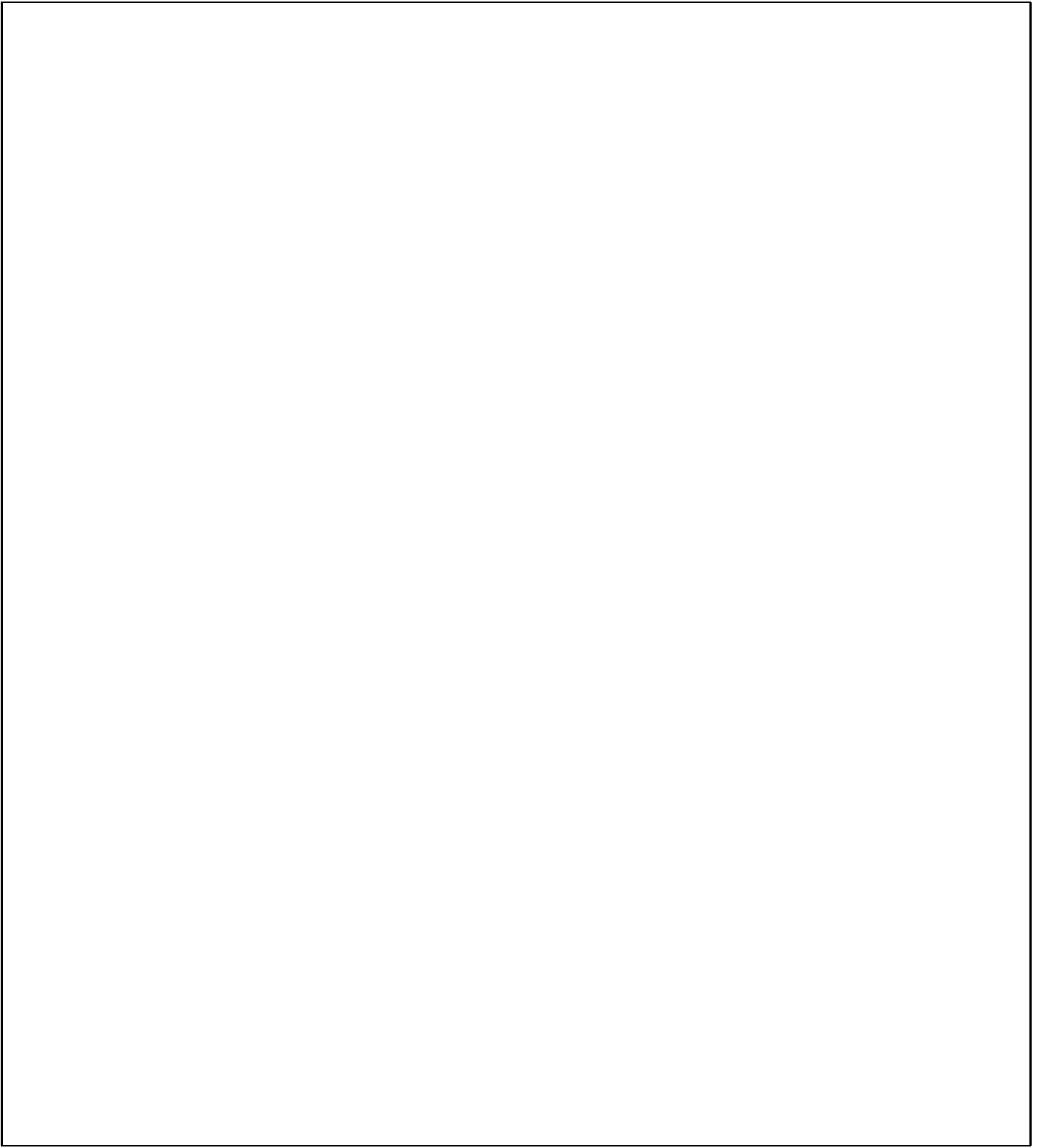

%\ \\
%\\[-1.0cm]
%\hspace*{-2.0cm}
%\psfig{figure=fig7.ps,width=20cm}
%\\[-5.25cm]
\picplace{20cm}
\caption[TEMPERATURE_A]{{\footnotesize{The time development of the corresponding temperature.}}}
\end{figure}
\twocolumn
\onecolumn
\begin{figure}[ht]
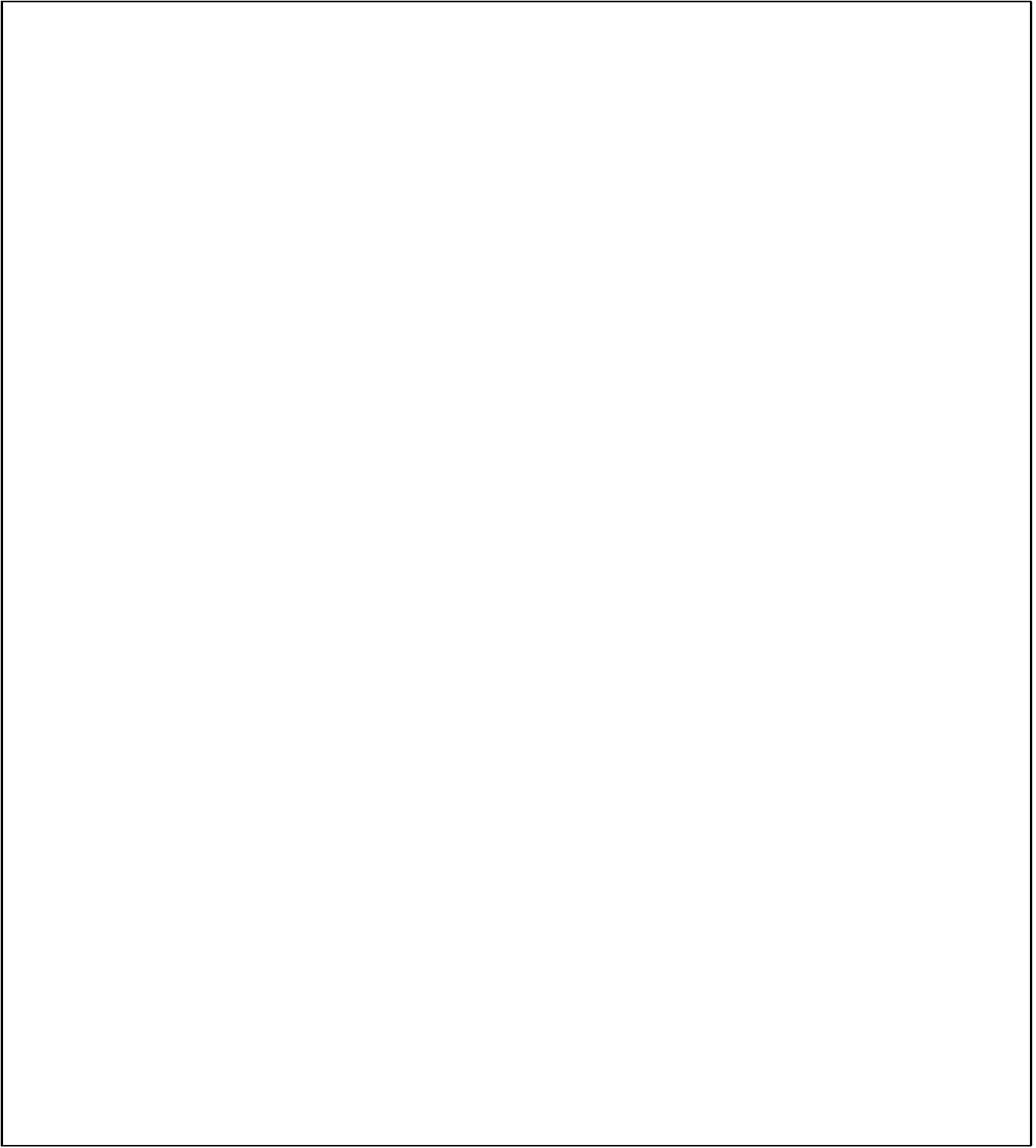

%\ \\
%\\[-1.0cm]
%\hspace*{-2.0cm}
%\psfig{figure=fig8.ps,width=20cm}
%\\[-5.25cm]
\picplace{20cm}
\caption[TEMPERATURE_B]{{\footnotesize{The time development of the temperature in the ``low temperature case''.}}}
\end{figure}
\twocolumn
The temperature reaches the observed temperatures of a few million
K degree within a very short time of 1.5 hours (cf. Fig. 7).
Two effects can easily be seen:
\begin{itemize}
\item The magnetized plasma is compressed, i.e.
mass density and temperature are increased.
\item
The temperature peak is the result of magnetic reconnection.
\end{itemize}
Both, magnetic compression and magnetic reconnection, lead to the observed temperatures.
We note that the plasma flows escaping from the reconnection zones mediate the
high temperature information to the surrounding medium.
A chain of processes
takes place in the plasma around the reconnection zones. First the kinetic energy of the
plasma flows is converted into heat and second the plasma flows induce
plasma instabilities in the gas. They result in an increase of the collision
frequency and a reduction of the electrical conductivity. Additionally,
magnetic reconnection leads to a relaxation of the agitated magnetic field.

\ \\
We started these simulations at nearly one million K degree.
However, since the Reynolds layer is much cooler than \mbox{900\,000 K },
we consider also the time
development of the temperature for the case, where the initial temperature is about
250 K and the initial density is about 22.5 particles per cm$^3$ (Fig. 8). 
The magnetic field strength
is again chosen as 17 $\mu$G and the Alfv\'en time is about 35 sec.
Temperatures of
more than 2000 K are easily reached within a very short time. Both effects, magnetic compression and
magnetic reconnection are clearly visible.

\ \\
\section{Cooling processes}

The simulations confirm the qualitative estimates that the plasma can be heated up to a few million K within a very short time.
To fill up a large volume with hot gas via magnetic heating the heating process must not
be overcome by radiative and thermal conductive losses, the cooling time $ t_{cool} $ and the conductive time $ t_{cond} $
have to be larger than the corresponding heating time $ t_{heat} $; 
if the gas would cool very quickly, magnetic heating could not explain 
the observations.

The conversion of kinetic energy into luminosity by collision processes cools a hot, partly or totally ionized
gas to lower temperatures. The principal cooling mechanism for a very hot plasma with temperature
more than 10$^7$ K is thermal bremsstrahlung or free-free emission. Gas at lower temperatures cools
mainly by electron impact excitation of electronic levels of the neutral and ionized particles.
Finally the fine structure excitation becomes more and more important. Taking all these
processes into account, Dalgarno \& McCray (1972) derived the interstellar
cooling function ${\cal{L}}$, shown in the following figure:

%\ \\
\begin{figure}[ht]
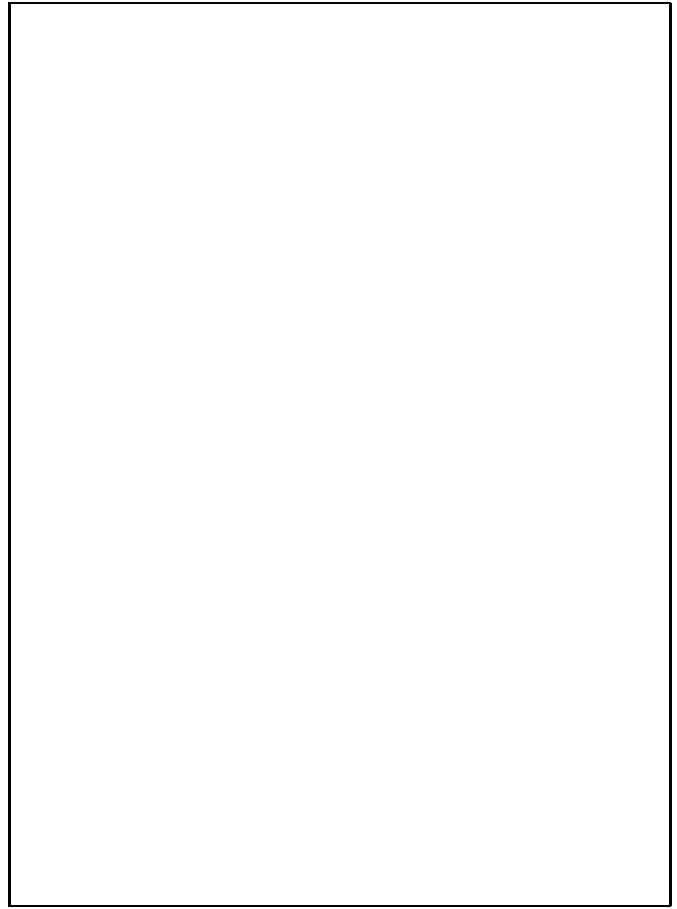

%\hspace*{2cm}

%\ \\
%\\[-3cm]
%\hspace*{-2cm}
%\psfig{figure=fig9.ps,width=12cm}
%\vspace*{1.5cm}
%\\[-6.5cm]
\picplace{12cm}
\caption[COOLING]{{\footnotesize{The interstellar cooling function ${\cal{L}}$ for various values of ionisation.}}}
\end{figure}

%\ \\
The cooling rate $\Lambda$ is given by the product of the square of the number density and the
interstellar cooling function ${\cal{L}}$:
\[ \Lambda = n^2\, {\cal{L}} \]
The cooling rate depends on the number density $n$, the temperature $T$ and the degree of ionization.
The ratio between the cooling time and heating time can be expressed in terms of the heating rate $Q$ and
the cooling rate $ \Lambda $:
\[ \frac{t_{cool}}{t_{heat}} = \left(\frac{n k_B T}{\Lambda}\right)\Big/\left(\frac{n k_B T}{Q}\right) = \frac{Q}{\Lambda} \]
The heating rate $ Q = \eta\,\vec{j}^2 $ due to magnetic reconnection is taken from our simulations.
Even if we use the maximum cooling rate of
\[ \Lambda_{max} = n^2\,{\cal L}_{max} = n^2 \cdot 10^{-21} \left[\frac{erg\,cm^3}{s}\right] \]
which is only valid for temperatures of 10$^{4-5}$ K, the cooling is never efficient
enough to restrict the magnetic heating in the whole temperature range we have in mind ($T_i$ = 10$^{2-6}$ K).
This can be seen from Fig. 10, which shows the ratio $\frac{Q}{\Lambda_{max}}$ of the heating
rate $Q$ to the maximum cooling rate $ \Lambda_{max}$ versus the initial temperature $T_i$:
In the ``low temperature-regime'' with an initial temperature of a few hundred K this ratio is about
10$^3$, i.e. the cooling time exceeds the corresponding heating time at least by a factor 10$^3$.
In a real plasma at low temperatures the degree of ionization is considerably smaller than unity, so the value
of the cooling function for these temperatures is overestimated by several orders of magnitudes. Therefore the
ratio between the cooling time and the heating time will be much larger than the value above.
In the ``high temperature case'' with an initial temperature of 9$\cdot$10$^5$ K this ratio is even
greater: $ \frac{Q}{\Lambda_{max}} \approx 10^{10} $ or $ t_{cool} \approx  10^{10} \, t_{heat} ! $
So, the cooling time is considerably greater than the corresponding heating time !

\ \\
\begin{figure}[ht]
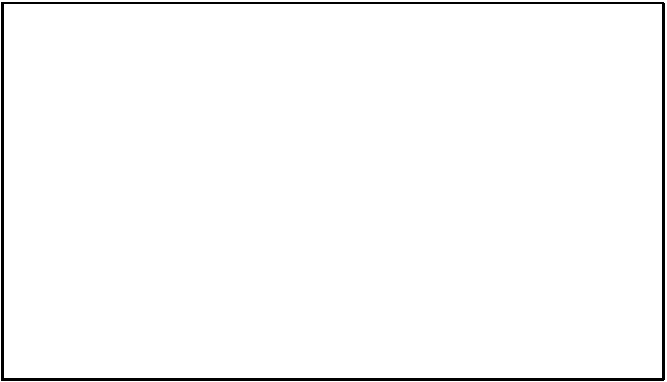

%\ \\
%\hspace*{0.25cm}
%\psfig{figure=fig10.ps,width=5cm,angle=-90}
%\vspace*{1.5cm}
%\\[-2.5cm]
\picplace{5cm}
\caption[HEATINGRATE]{{\footnotesize{The ratio between the heating rate Q and the maximum cooling rate $ \Lambda_{max} $ in dependence
of the initial temperature $T_i$.}}}
\end{figure}

To construct a simple analytic expression of the heating rate,
we use a plane sheet with length $L$,
surface $L^2$ and the thickness $l$.

The thickness $l$ of a reconnection sheet has been determined by
Parker (1979, p. 392): In the steady state Ohmic dissipation across the sheet is
$l j^2/\sigma$ is just sufficient to devour the influx of magnetic energy
$w B_{\rm 0}^2/{8\pi}$ from either side where $w$ is the velocity with which two
opposite fields move steadily towards each other.
The total pressure $p+{B_{\rm 0}\over{8\pi}}$ is uniform across the
sheet. The gas pressure $p$ attains its highest value where $B$ goes to zero.
This pressure excess ejects plasma from the opposite fields, along the field
lines. The velocity of expulsion is just equal to the Alfv\'en speed $v_{\rm A}$.
Conservation of fluid mass requires that the net magnetic field
inflow balances the outflow
\begin{equation}
wL=v_{\rm A} l.
\end{equation}
where $L$ denotes the width of the quasistationary current sheet.
In terms of the magnetic Reynolds number $R_{\rm M}$
\begin{equation}
R_{\rm M}={2L\,v_{\rm A}\over{\eta{\rm _R}}},
\end{equation}
($\eta_{\rm R}=c^2/(4\pi\sigma)$ denotes the resistive diffusion coefficient)
and with
\begin{equation}
l{j^2\over\sigma}=w{B^2\over{8\pi}},
\end{equation}
one obtains
\begin{equation}
l={2L\over{\sqrt{R_{\rm M}}}},
\end{equation}
and
\begin{equation}
w={2v_{\rm A}\over{\sqrt{R_{\rm M}}}}.
\end{equation}
Inserting $w$ and $l$ into Eq. (11) and using $Q=j^2/\sigma$ we end up with

\begin{equation}
Q={B^2\over{8\pi}}{v_{\rm A}\over L}.
\end{equation}

By equating $Q$ with the maximum value of the cooling 
function ${\cal L}_{max}$, we can determine
what length scales can be heated up to the maximum value until cooling
restricts the heating:

\begin{equation}
L\sim {B^2\over{8\pi}}{v_{\rm A}\over{10^{-21} n^2}}
\end{equation}

\ \\
\begin{figure}[ht]
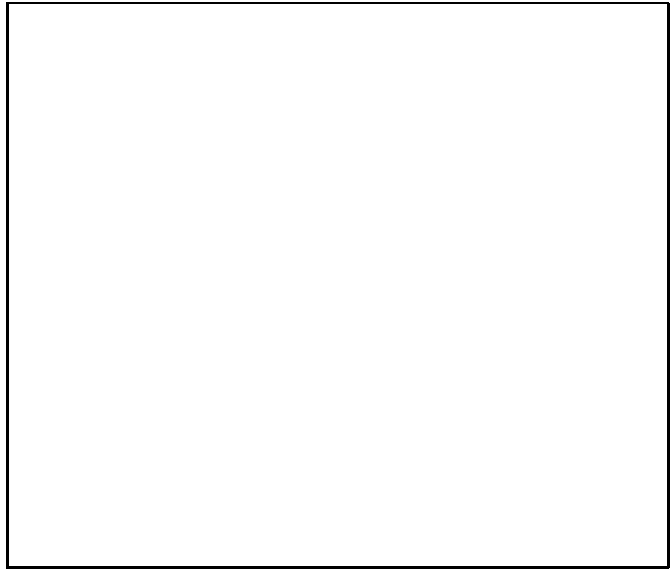

%\ \\
%\\[1.0cm]
%\hspace*{0.5cm}
%\psfig{figure=fig11.ps,width=5cm,angle=-90}
%\vspace*{1.5cm}
%\\[-2.5cm]
\picplace{7.5cm}
\caption[LENGTH]{{\footnotesize{$L$ in dependence of the number density $n$}}}
\end{figure}

Fig. 11 shows $L$ for the density we have in mind. Obviously the heating is
much more efficient than even the strongest cooling agent for the spatial
scales we need to explain the X-ray temperatures connected with impacting HVCs
$\sim$ pc. Only very dense regions with densities of about $10$ cm$^{-3}$ can be
influenced by cooling, but there we overestimate the cooling function by
several orders of magnitudes, as can be seen from Fig. 9, since the high
density regions are only partially ionized because of their small initial temperatures
($\sim$ 100 K).

\ \\
We know that magnetic reconnection is more effective in small localized regions, i.e. the thermal
energy has to be transported by thermal conductive processes. During the interaction
of a HVC with the galactic halo the gas will be heated up. Since the observations with ROSAT
show that the X-ray emitting hot gas is restricted to an enclosed area, the
typical time scale for thermal conductive processes has to be much larger than the
heating time. The physics of heat transport in magnetized plasmas is still a matter of debate.
Especially the influence of magnetic fields on the thermal conductivity $ \kappa $ is still unclear.
The widespread believe that tangled magnetic fields reduce $ \kappa $ by several orders of
magnitudes was shown to be highly questionable (Rosner \& Tucker 1989).
We therefore use the classical value of the thermal conductivity for a hydrogen plasma (Spitzer 1956):
\[ \kappa \simeq  1.31\,n_e\,\lambda_e\,k_B\,\left(\frac{k_B T_e}{m_e} \right)^{1/2}           \]
with
\[ \lambda_e = \frac{3^{3/2}\,\left(k_B T_e \right)^2}{4\,\pi^{1/2}\,n_e\,e^4\,ln\Lambda} \]
$T_e$ is the electron temperature, $n_e$ the number density and 
$ln\,\Lambda$ denotes the Coulomb logarithm, i.e.
the ratio of largest to smallest impact parameters for the collisions.
The conductivity time scale is approximately given by
\[ t_{cond} \simeq \frac{n L_{\nabla T}^2 k_B}{\kappa} \]
where $ L_{\nabla T} := \frac{T}{| \nabla T |} $ is the length scale on which the temperature changes. So we find for the
ratio between the conductivity time scale and the heating time:
\begin{eqnarray*}
\frac{t_{cond}}{t_{heat}} &\simeq& \frac{4 \pi^{1/2}\,e^4\,ln\Lambda\,m_e^{1/2}}{1.31\,\cdot\,3^{3/2}\,k_B^{7/2}} \, \frac{L_{\nabla T}^2\,Q}{T_e^{7/2}} \nonumber \\
                          &\simeq& 2.1585 \cdot 10^6 \, \left[ \frac{ln\Lambda}{40} \right]\, \frac{L_{\nabla T}^2\,Q}{T_e^{7/2}}
\end{eqnarray*}
A ratio of about 10$^3$ between these time scales requires $L_{\nabla T} \le 5 \cdot 10^{-8} $ pc for the
low temperature gas and $L_{\nabla T} \le 5 \cdot 10^{-2} $ pc for the high temperature regime ! Since the observed
length scales are much larger, $t_{cond}$ will exceed the heating time by a factor of more than 10$^3$.
Kerp \& G{\"u}sten (1996, in prep.) detected a magnetic field located close to the X-ray brightest part
of HVC 90.5+42.5-130 via HI 21cm Zeeman observations with the Effelsberg telescope with a linear size
of about 10 pc. Such fields are well ordered and will substantially reduce the thermal conductivity
perpendicular to the field lines. Thus, the hot gas in the boundary between HVC and halo
is thermally well isolated from the halo.

\section{Summary}

We have investigated whether compressional heating of magnetized plasmas
and magnetic reconnection can play important roles in  
the heating of cosmic plasmas. As an example we used
the impact of a magnetized high-velocity cloud onto the magnetized galactic
halo and the observed correlation with X-ray emission. The clouds move with
velocities of about 100 km\,s$^{-1}$ and heat a gas to higher temperatures
then kinetically expectable (some $10^5$ K!).
Thus, we idealize the interaction of the cloud with the halo by a very general
physical situation of two magnetized plasmas
which encounter with a velocity corresponding to the observed HVC-speeds.
Due to the high conductivity of both plasma components the first stage of the
interaction is dominated by compression of the field lines perpendicular to
the direction of motion of the cloud. This compression alone already increases
the pressure and temperature of the material in the boundary layer between
the cloud and halo. Due to internal motions inside the layer, the compressed field
lines are stretched, twisted, strained and curled by plasma motions. Thereby
magnetic field lines with antiparallel directions encounter. The subsequent
chain of electrodynamic processes is known as magnetic reconnection and corresponds
to a strongly enhanced dissipation of the kinetic energy stored in the compressed
magnetic fields via Ohmic heating $\propto j^2/\sigma$. The interaction of two highly conducting
plasmas provides both necessary conditions for enhanced Ohmic heating: first it
leads to an increase of the current density $j$ by compression and second if a critical
current density is exceeded
plasma instabilities are excited, which increase the collision frequency by many orders of
magnitude, thereby reducing the electrical conductivity $\sigma$ significantly.

We have shown that magnetic heating in low density plasmas is not restricted
by even the most effective radiative cooling or heat conduction. The agitated
plasma will always reach the maximal attainable temperature, deduced from
the magnetic pressure $B^2/8\pi$. Our model explains the overall energetics and
the observed temperatures.

Since both heating mechanisms - magnetic compression and magnetic reconnection - are
unavoidable in magnetized
plasmas which are agitated by randomly directed gas motions,
we think that the dissipation of magnetic fields in cosmic plasmas presents
a new possibility in the context of astrophysical heating and cooling
processes. Since most of the astrophysical plasmas can be envisaged as
highly conductive, any plasma motion is accompanied by the motion of magnetic field
lines. Any velocity component perpendicular to the field lines compress the magnetic field.
This compression increases the magnetic pressure faster than the thermal
pressure is enhanced, i.e. $B\propto \rho$, $B^2\propto \rho^2$ and for isotropic
compression $B^2\propto \rho^{4/3}$. In any case the magnetic pressure becomes
more and more important during the interaction of streaming plasmas. Since the
magnetic pressure has its physical origin in the Lorentz-force density
$\propto {\bf j \times B}$, an enhancement of the magnetic pressure
is synonymous with an increase of the current density and the dissipation rate.
The actual location where this dissipation takes place depends on the field
direction and the local electrical conductivity, which in turn depends on the current density!
This electrodynamic coupling
scenario allows the localized release of energy which has been stored in globally compressed
magnetic field lines.

\ \\
\ \\
{\sl Acknowledgements.} This work was  supported by the the 
Deutsche Forschungsgemeinschaft 
through the grant ME 745/18-1. 
F. Zimmer was supported by the Bennigsen-Foerder Award (1994) of the
government of Nordrhein-Westfalen.

{}

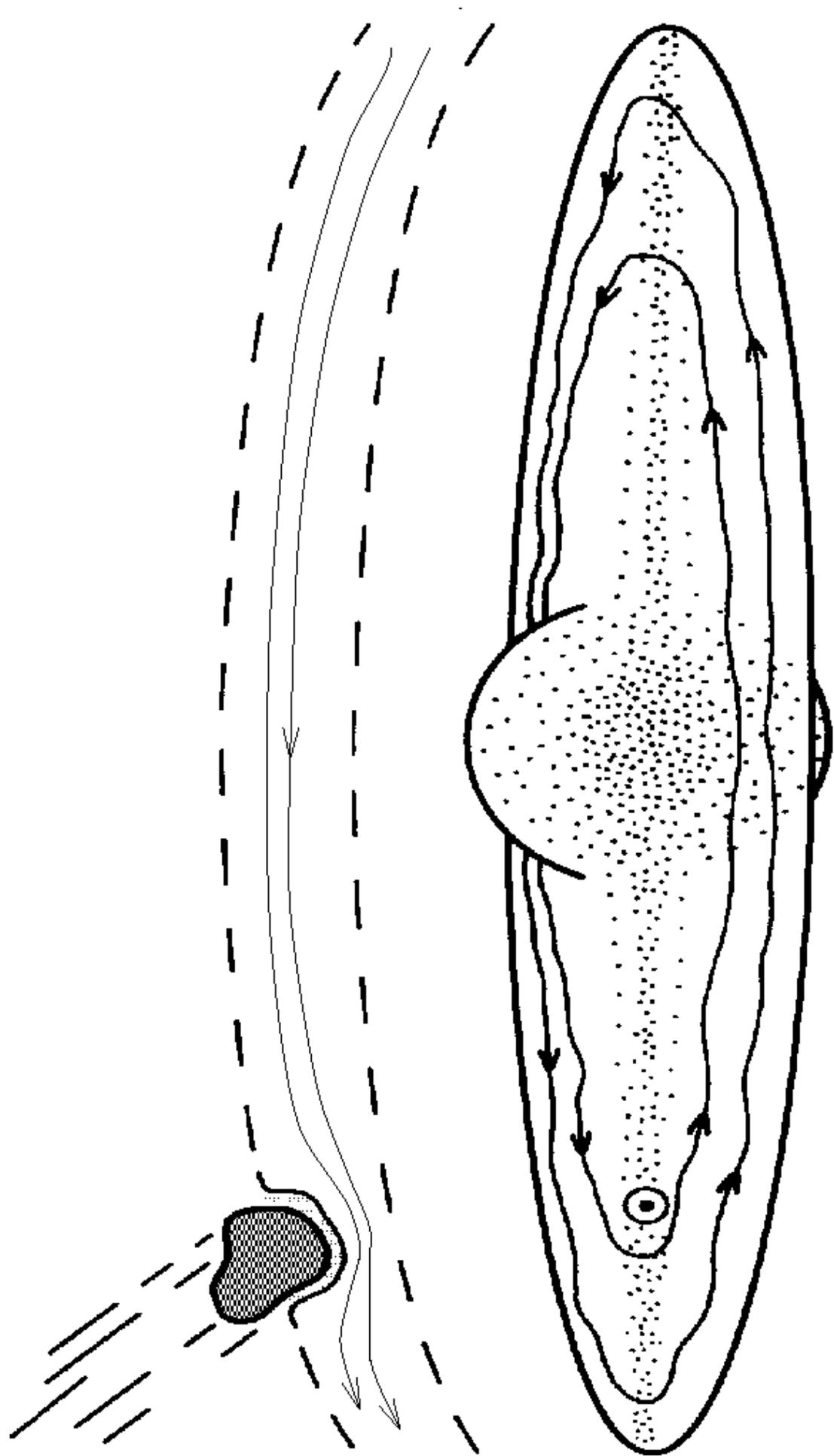
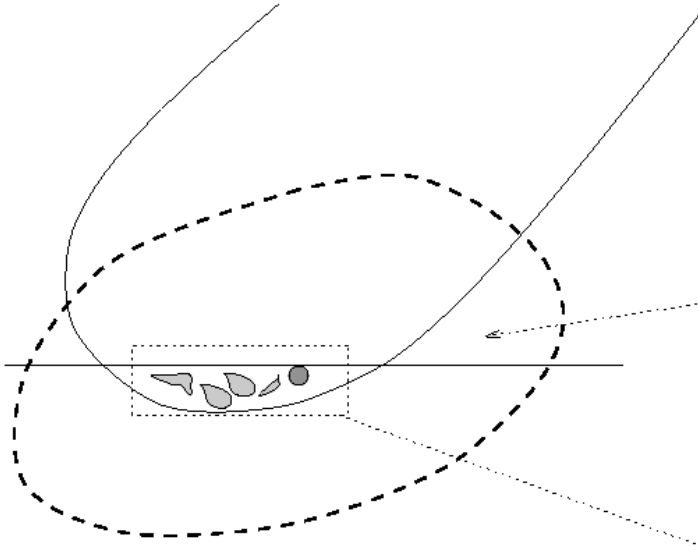
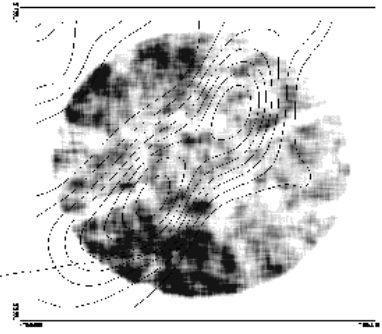
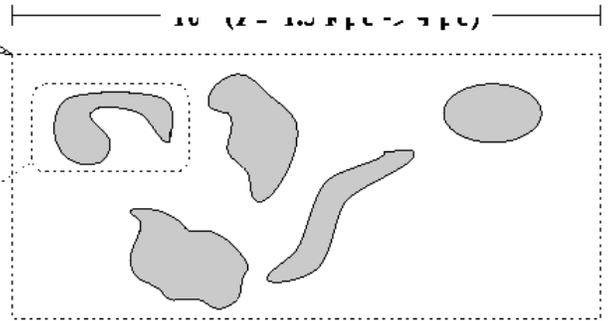
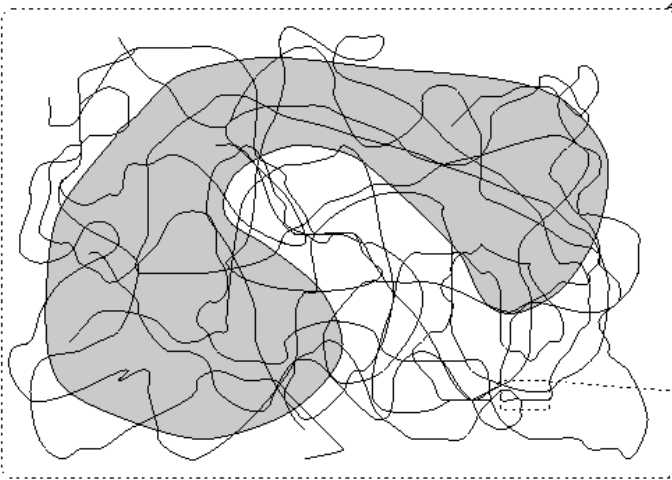
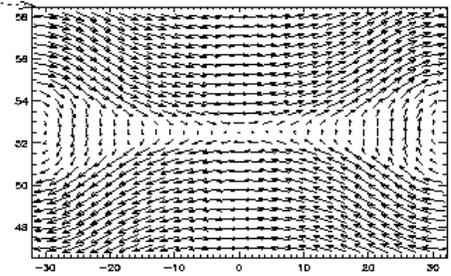
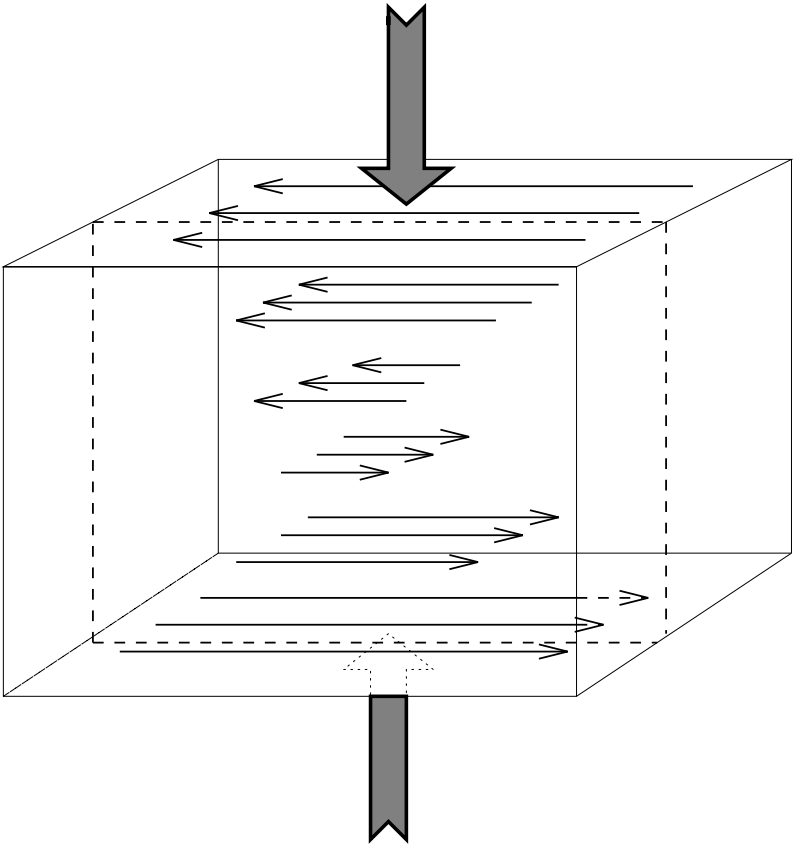
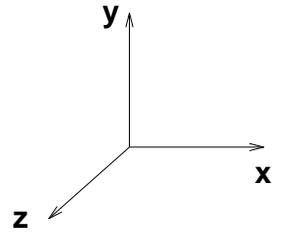
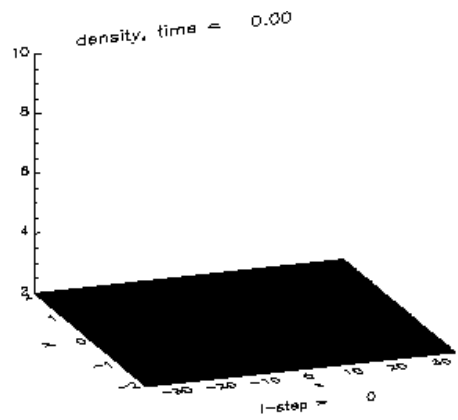
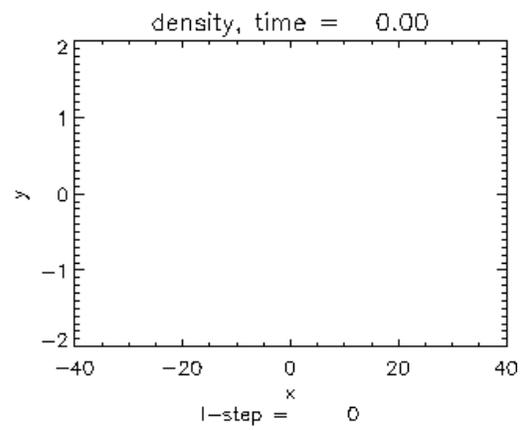
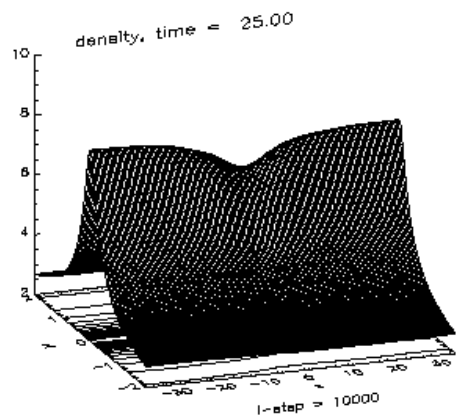
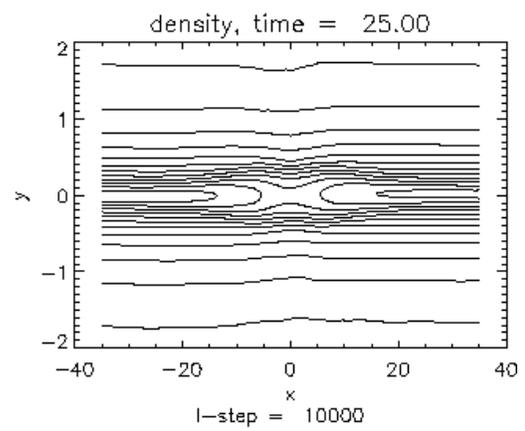
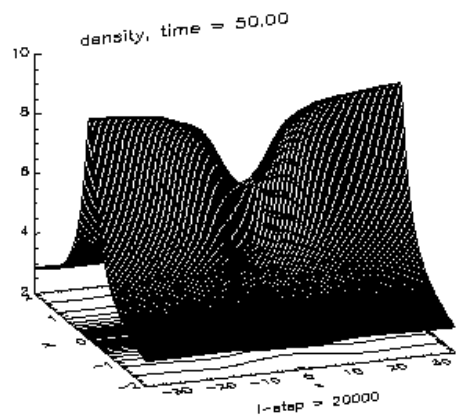
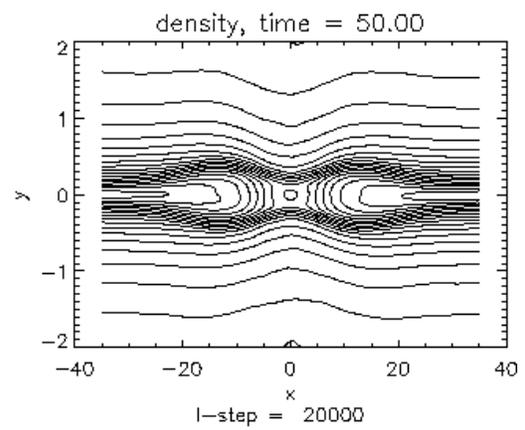
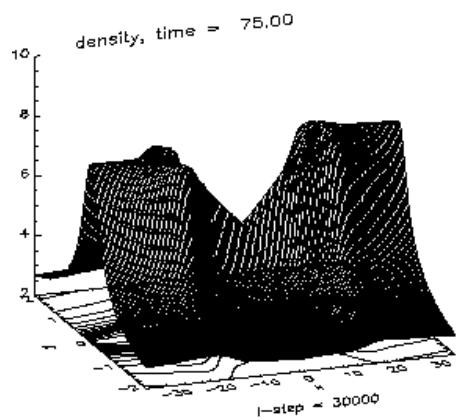
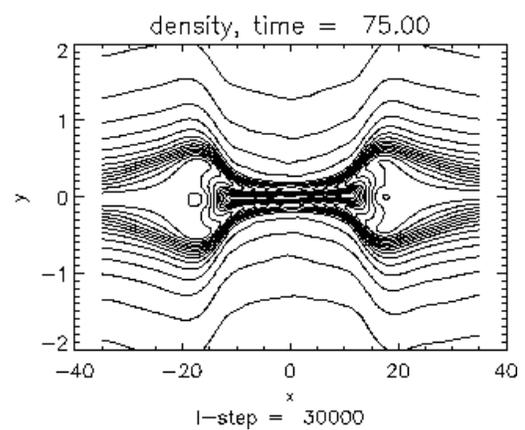
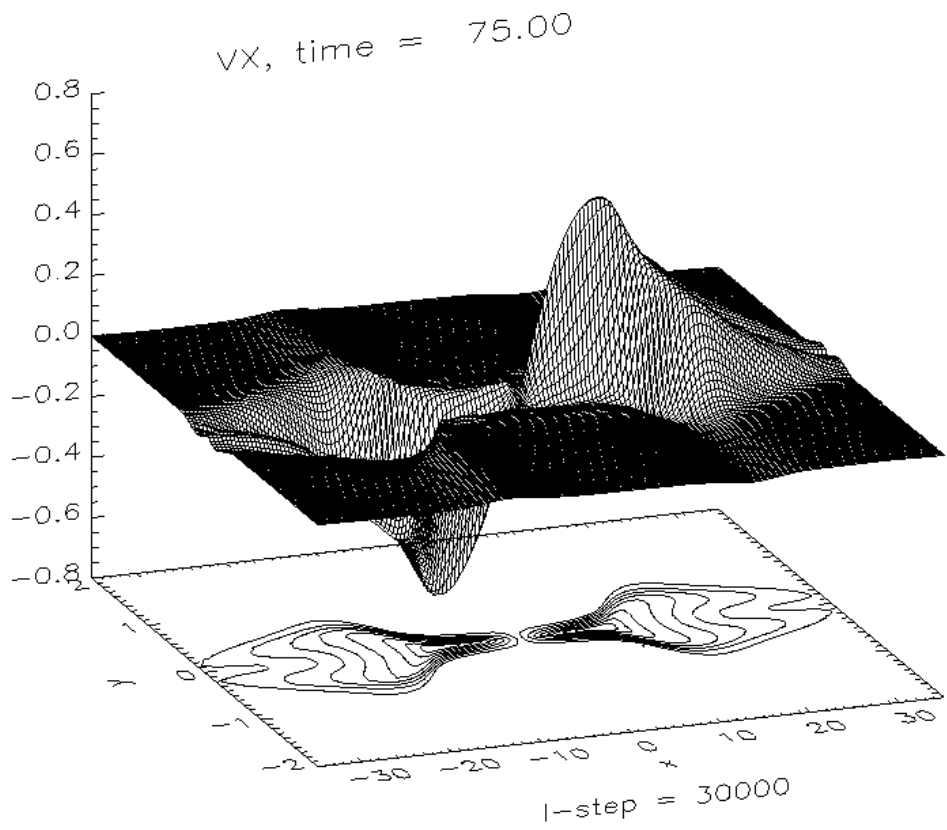
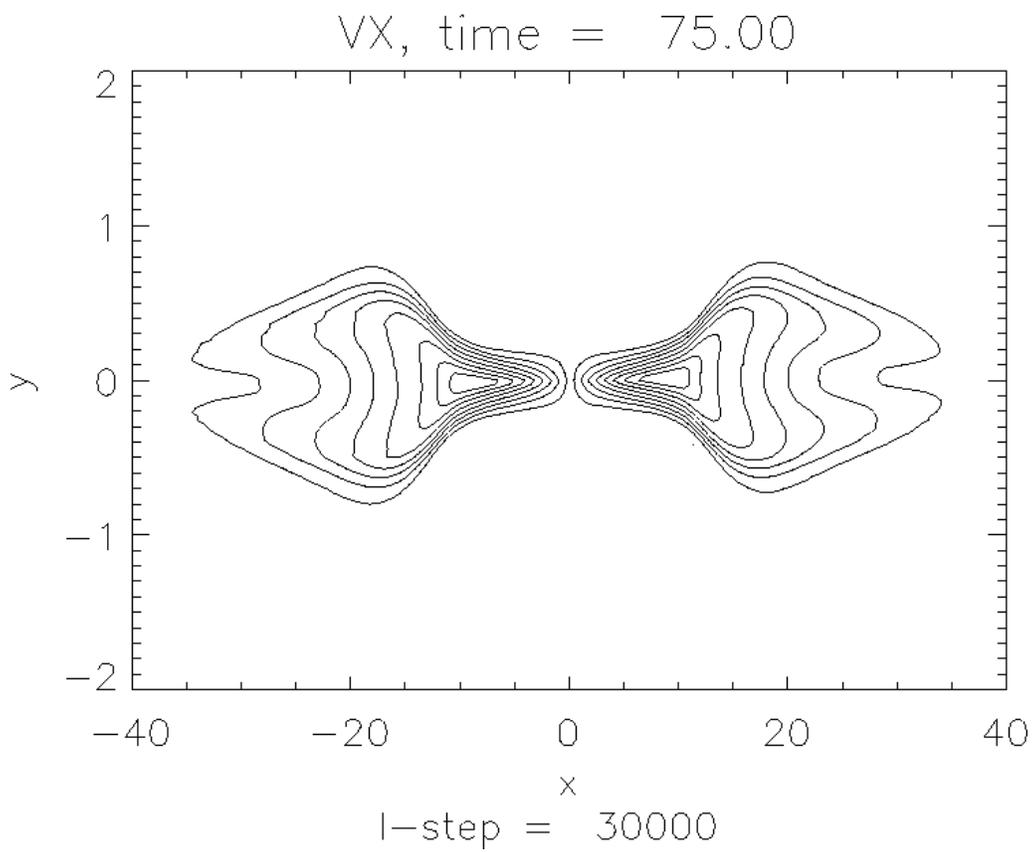
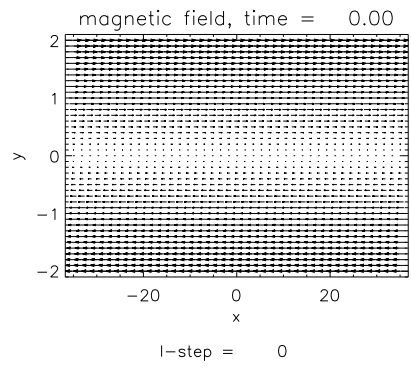
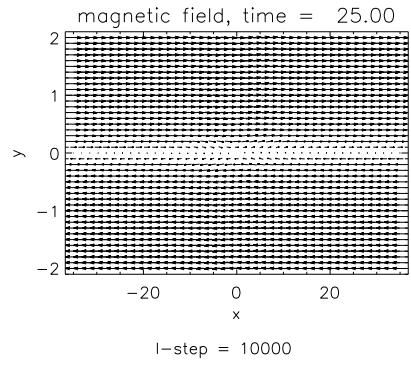
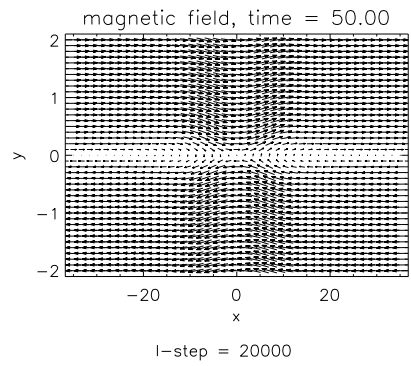
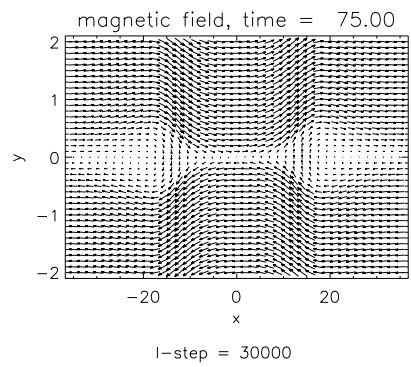
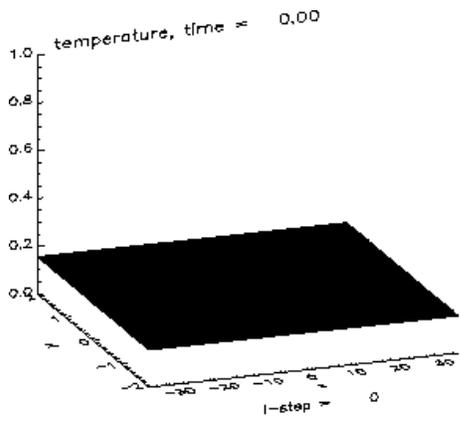
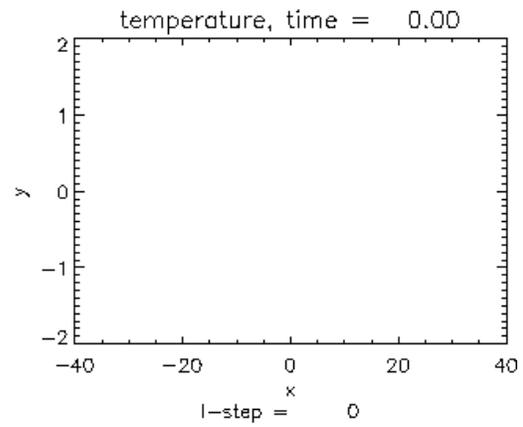
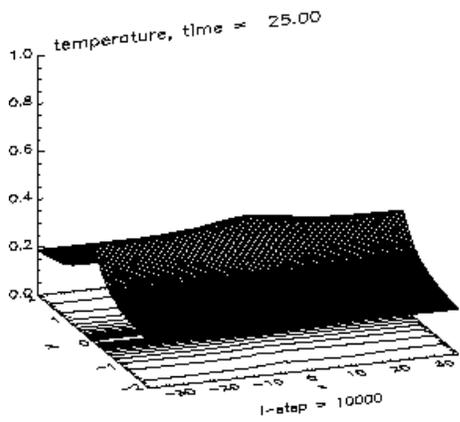
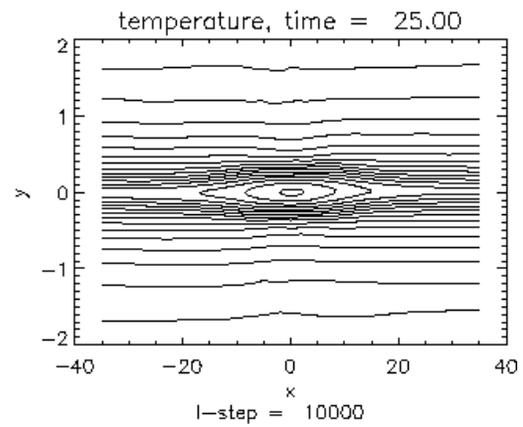
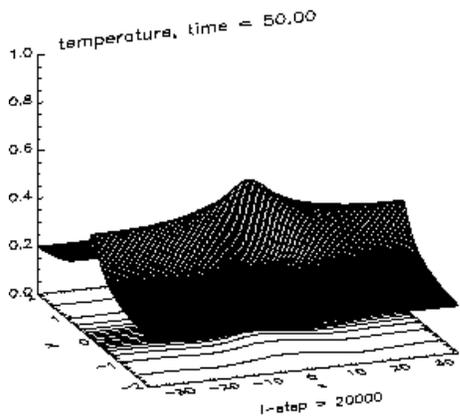
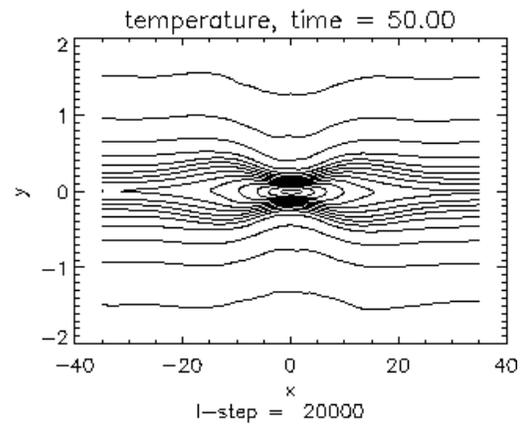
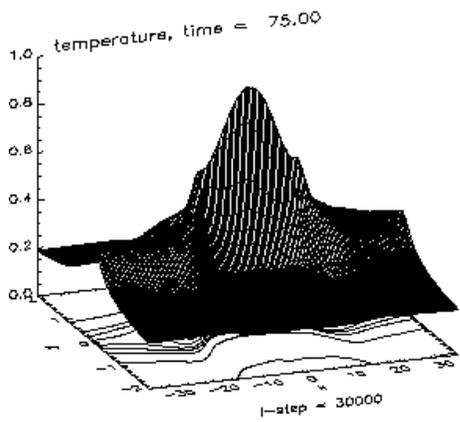
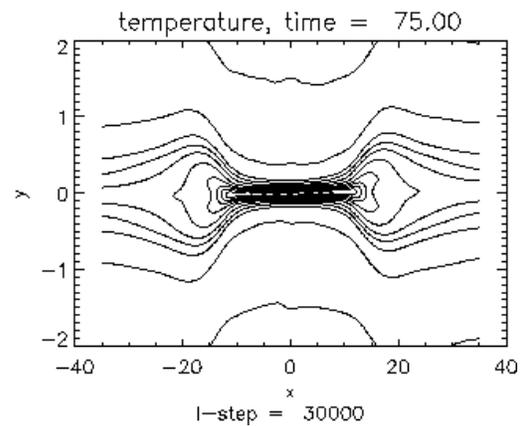
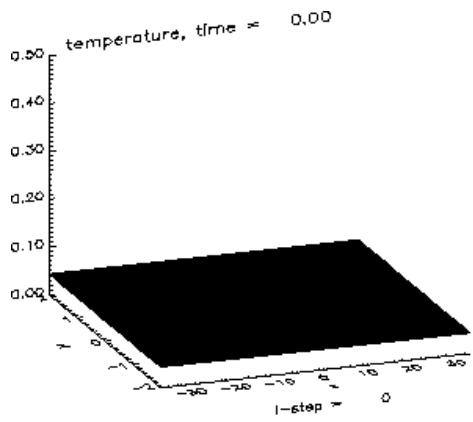
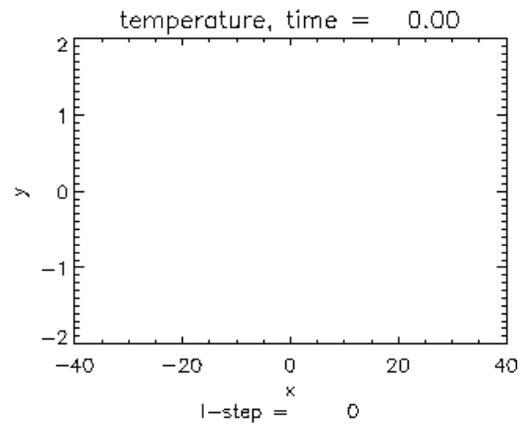
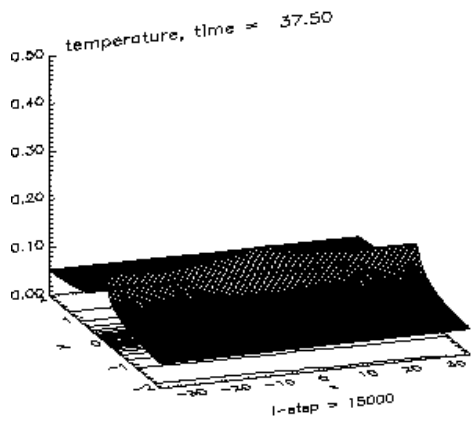
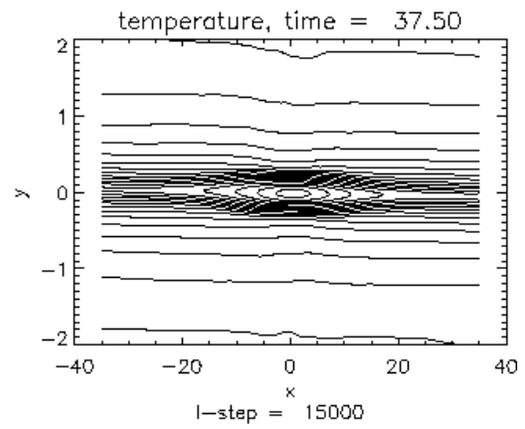
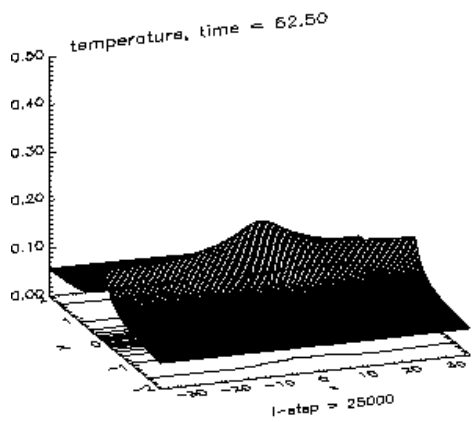
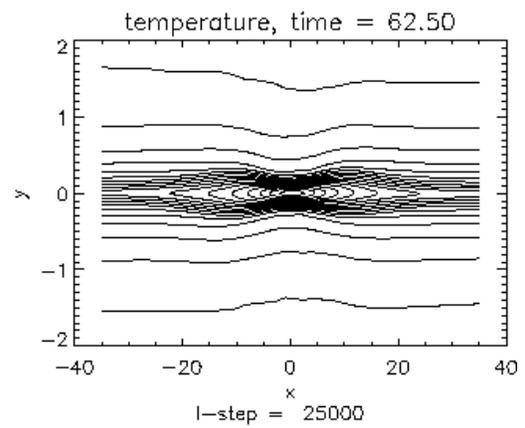
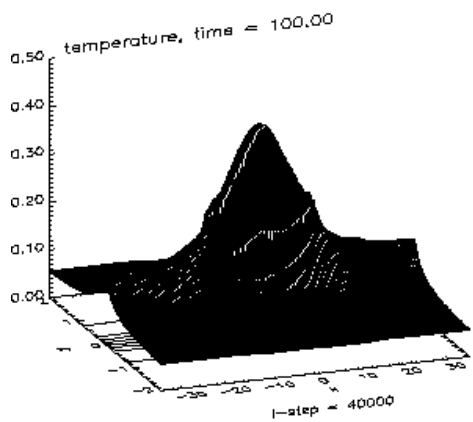
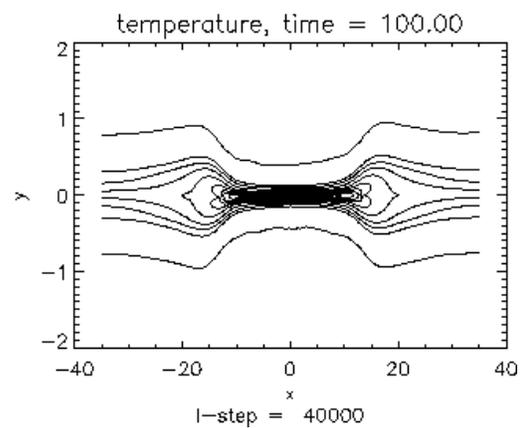
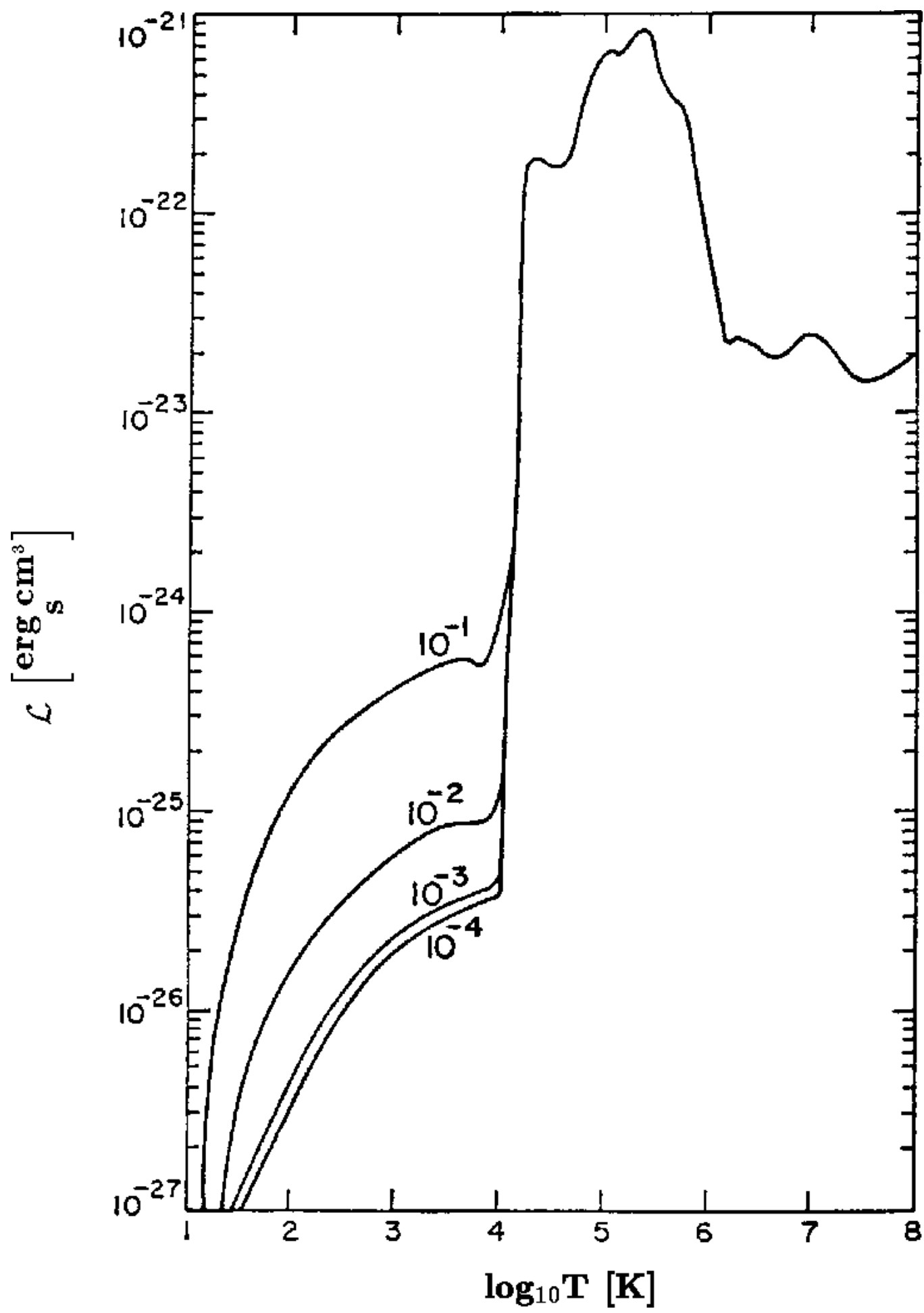
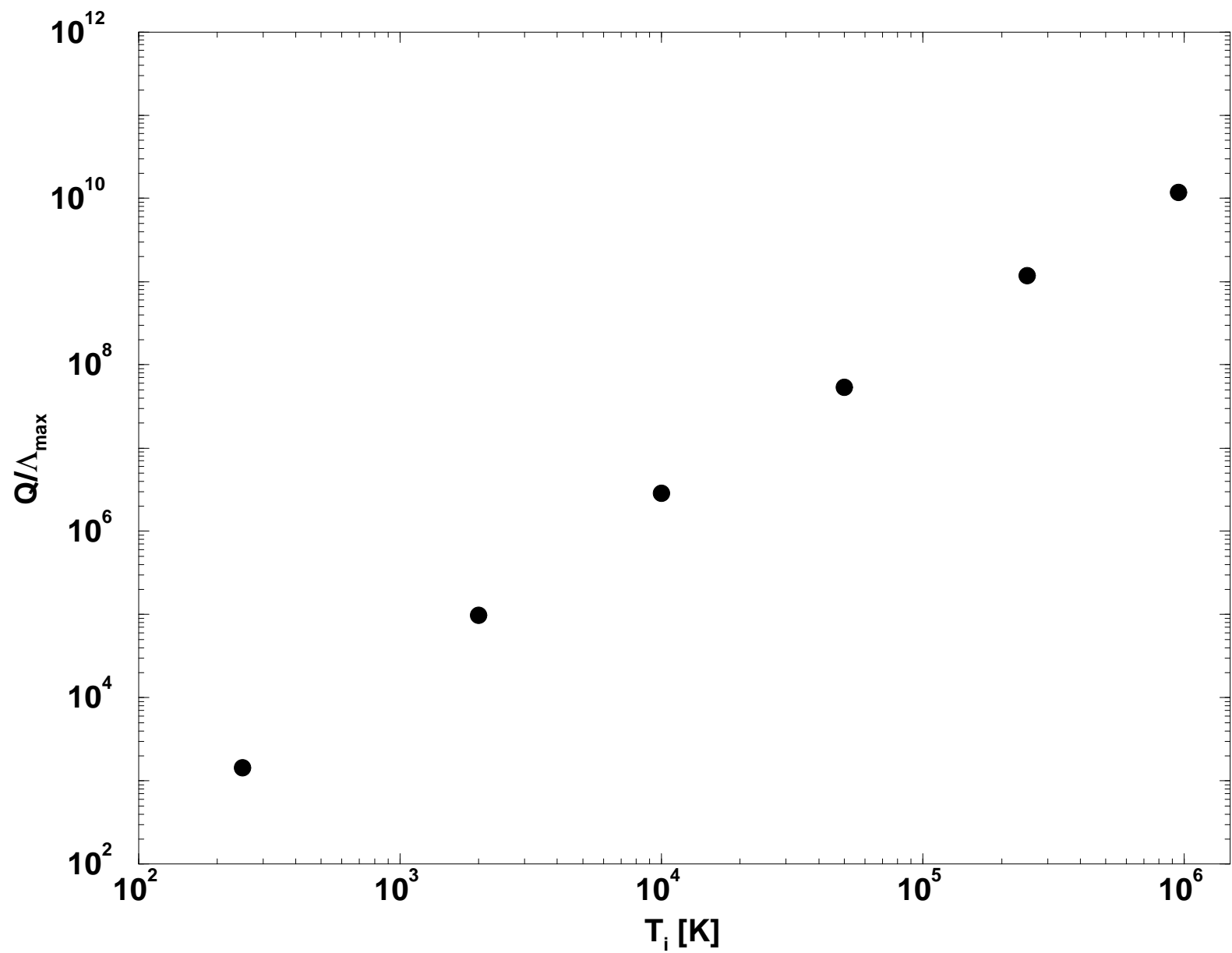
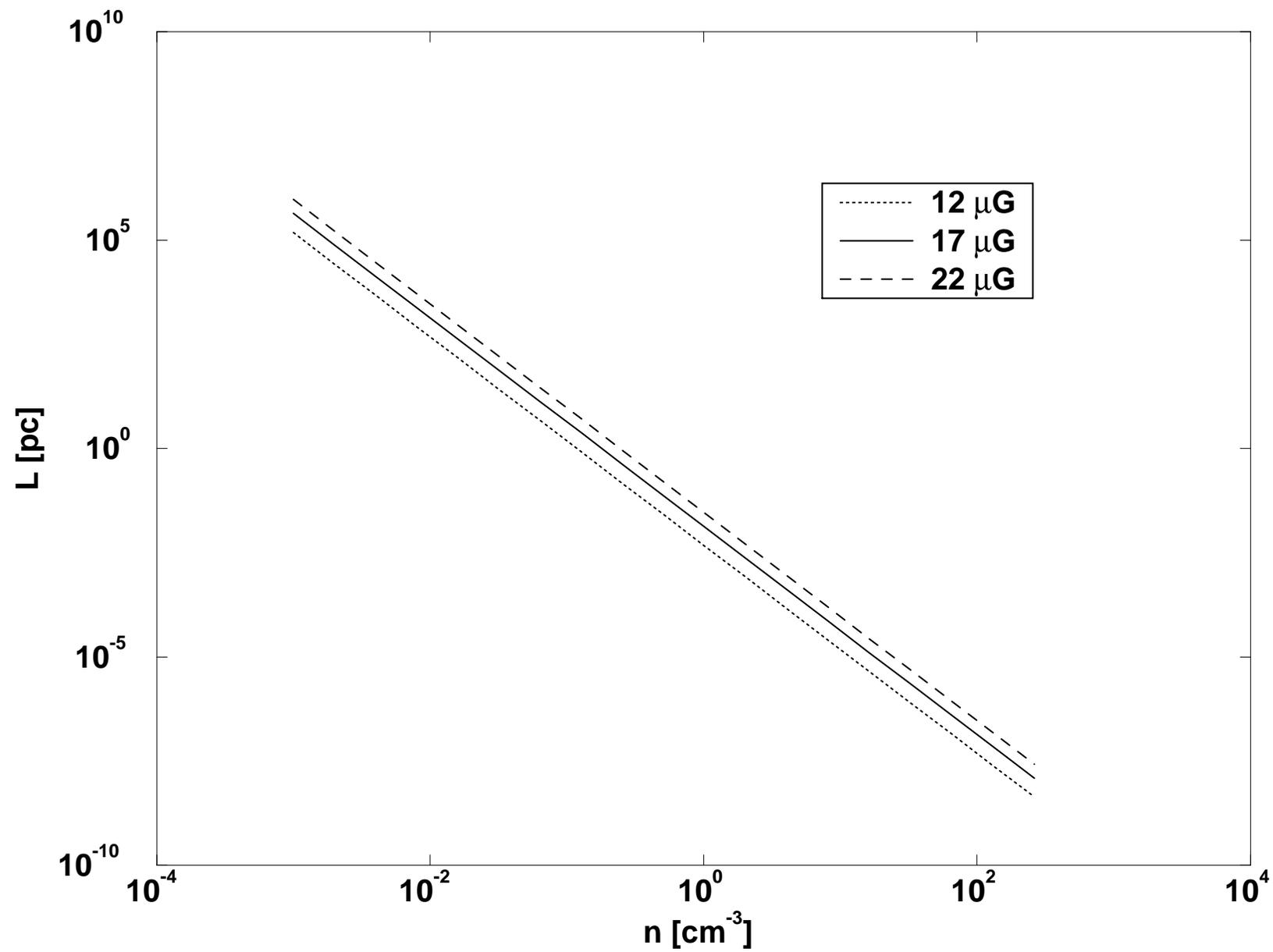
\end{document}